\documentstyle[sprocl]{article}

\input{psfig}

\def\be{\begin{equation}}
\def\ee{\end{equation}}
\def\bea{\begin{eqnarray}}
\def\eea{\end{eqnarray}}
\begin{document}

\title{QUARK ASYMMETRIES IN THE PROTON}

% Talk given at the
%	Workshop on Future Developments in Quark-Nuclear Physics,
%	Adelaide, March 1998

\author{W. MELNITCHOUK}

\address{Institut f\"ur Kernphysik,
	Forschungszentrum J\"ulich,
	D-52425 J\"ulich, Germany	\\
	{\tt w.melnitchouk@fz-juelich.de}}

\maketitle

\abstracts{We review recent experimental and theoretical developments
	in the study of the sea quark structure of the proton.
	In the light quark sector, we analyse the recent $pp$ and $pD$
	Drell-Yan data from the E866/NuSea experiment at Fermilab,
	and their implication on the $\bar d/\bar u$ asymmetry in the
	proton.
	The current status of the strange content of the proton,
	including the possible difference between strange and
	antistrange quark distributions and strangeness form factors,
	is updated.
	Finally, we point out the implications of the possible
	non-symmetric charm and anticharm distributions in the nucleon
	for HERA event rates at large $x$ and $Q^2$.}

%%%%%%%%%%%%%%%%%%%%%%%%%%%%%%%%%%%%%%%%%%%%%%%%%%%%%%%%%%%%%%%%%%%%%%%%
\section{Introduction}

The sea of the proton is extremely fertile ground for the study of
non-perturbative QCD dynamics and its relation to the substructure
of hadrons.
In particular, sea quark asymmetries almost universally signal
the presence of interesting non-perturbative phenomena, on the
background of a perturbative QCD landscape which in comparison
is relatively flat and homogeneous.

%%%%%%%%%%%%%%%%%%%%%%%%%%%%%%%%%%%%%%%%%%%%%%%%%%%%%%%%%%%%%%%%%%%%%%%%
\section{Light Antiquark Asymmetry}

The E866/NuSea Collaboration recently measured the spectrum of
$\mu^+\mu^-$ Drell-Yan pairs produced in $pp$ and $pD$ collisions
at the FNAL Tevatron \cite{E866}, which has for the first time
enabled the shape of the $\bar d / \bar u$ ratio to be mapped out
over a large range of $x$.
The relatively large asymmetry found implies the presence of
non-trivial dynamics in the $\bar u$ and $\bar d$ sector of the
proton sea which does not have a perturbative QCD origin.
The novel and unexpected feature of the E866 data is that the
$\bar d/\bar u$ asymmetry peaks at rather small values of $x$,
$x \sim 0.15$, and drops quite rapidly to unity by $x \sim 0.3$.

The simplest and most obvious source of a non-perturbative asymmetry
in the light quark sea is the chiral structure of QCD.
{}From numerous studies in low energy physics, including chiral
perturbation theory, pions are known to play a crucial role in the
structure and dynamics of the nucleon, and there is no reason to
believe that the long-range tail of the nucleon should not also play
a role at higher energies.
As pointed out by Thomas \cite{AWT83}, if the proton's wave function   
contains an explicit $\pi^+ n$ Fock state component, a deep-inelastic
probe scattering from the virtual $\pi^+$, which contains a valence  
$\bar d$ quark, will automatically lead to a $\bar d$ excess in the
proton.
This is the essential physical idea behind these expectations,
and has been used to address not only the $\bar d / \bar u$
asymmetry (see Ref.\cite{REVIEW} and references therein),
but also SU(3) flavour symmetry breaking in the proton sea
\cite{AWT83}, as well as asymmetries in the strange \cite{MM}
and heavier flavour sectors \cite{CHARM,PNNDB}, as discussed below.

The essential ingredients of the meson cloud model are the
meson--baryon distribution functions, $f_{MB}(y)$, which give
the probability to find a meson, $M$, in the nucleon carrying
a fraction $y$ of the nucleon's light-cone momentum.
In a hadronic basis the only parameters on which these splitting
functions depend are the hadronic vertex functions, or form factors
at the nucleon--meson--baryon vertices, which are characterised
by an effective momentum scale (cut-off) $\Lambda_{MB}$.
{}From previous studies we know that the dominant contribution comes
from the $\pi N$ component of the nucleon, though the $\pi \Delta$
configuration turns out to play a crucial role here also.
In Fig.1(a) we show the $\pi N$ and $\pi \Delta$ momentum
distributions for (dipole cut-offs)
$\Lambda_{\pi N} = 1$~GeV, $\Lambda_{\pi\Delta} = 1.5$~GeV.
The relative magnitudes of these are taken from a comparison of
the axial form factors for the nucleon and for the $N$--$\Delta$
transition, which strongly favours an $N$--$\Delta$ axial form factor
that is significantly harder than that of the nucleon.
The harder $\pi \Delta$ distribution at large $y$ turns out to be
quite important phenomenologically for understanding the E866 data.

\begin{figure}[ht]
\centering{
\begin{picture}(-30,150)(180,0)
\psfig{figure=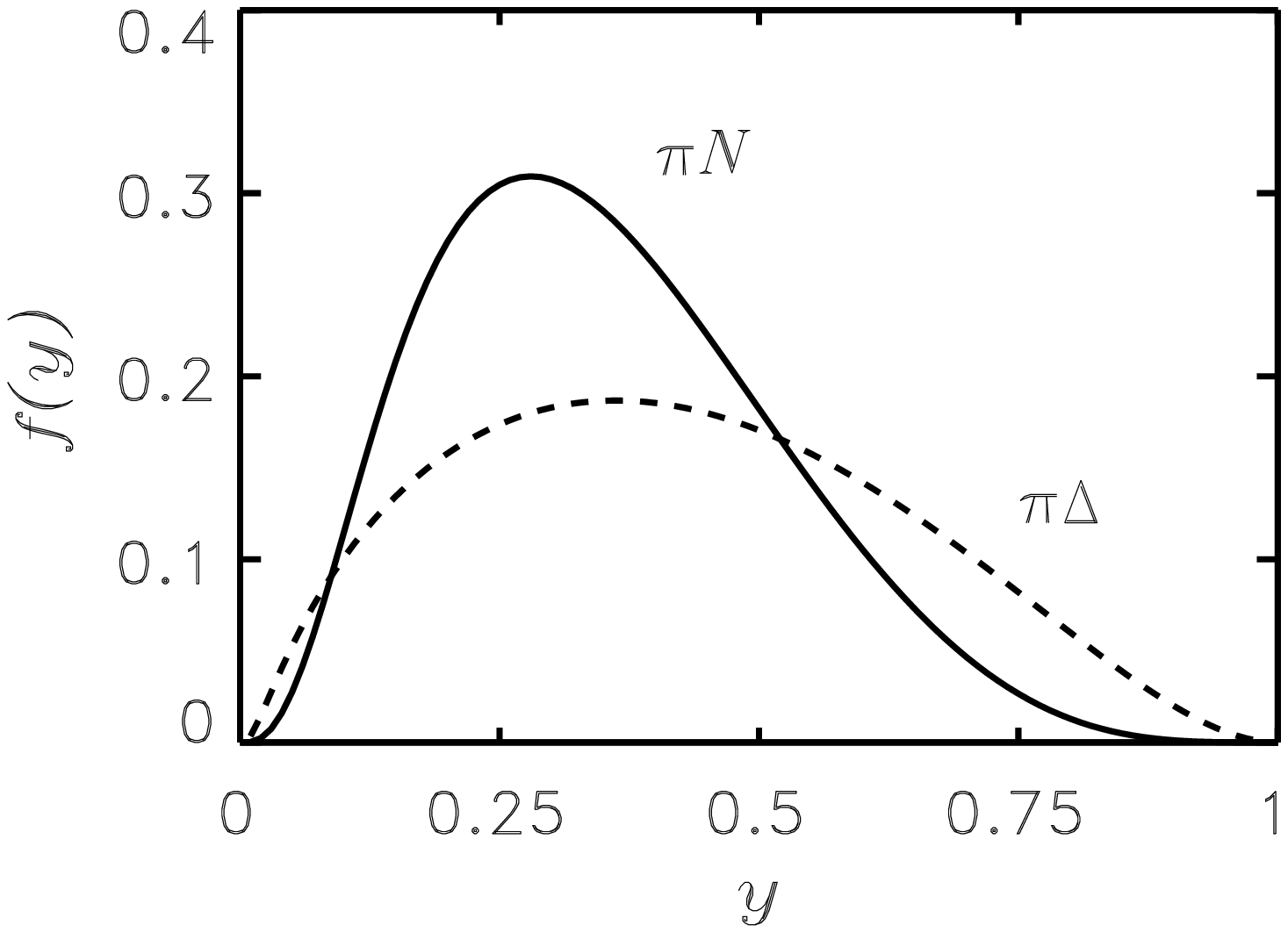,height=1.5in}
\put(0,0){\psfig{figure=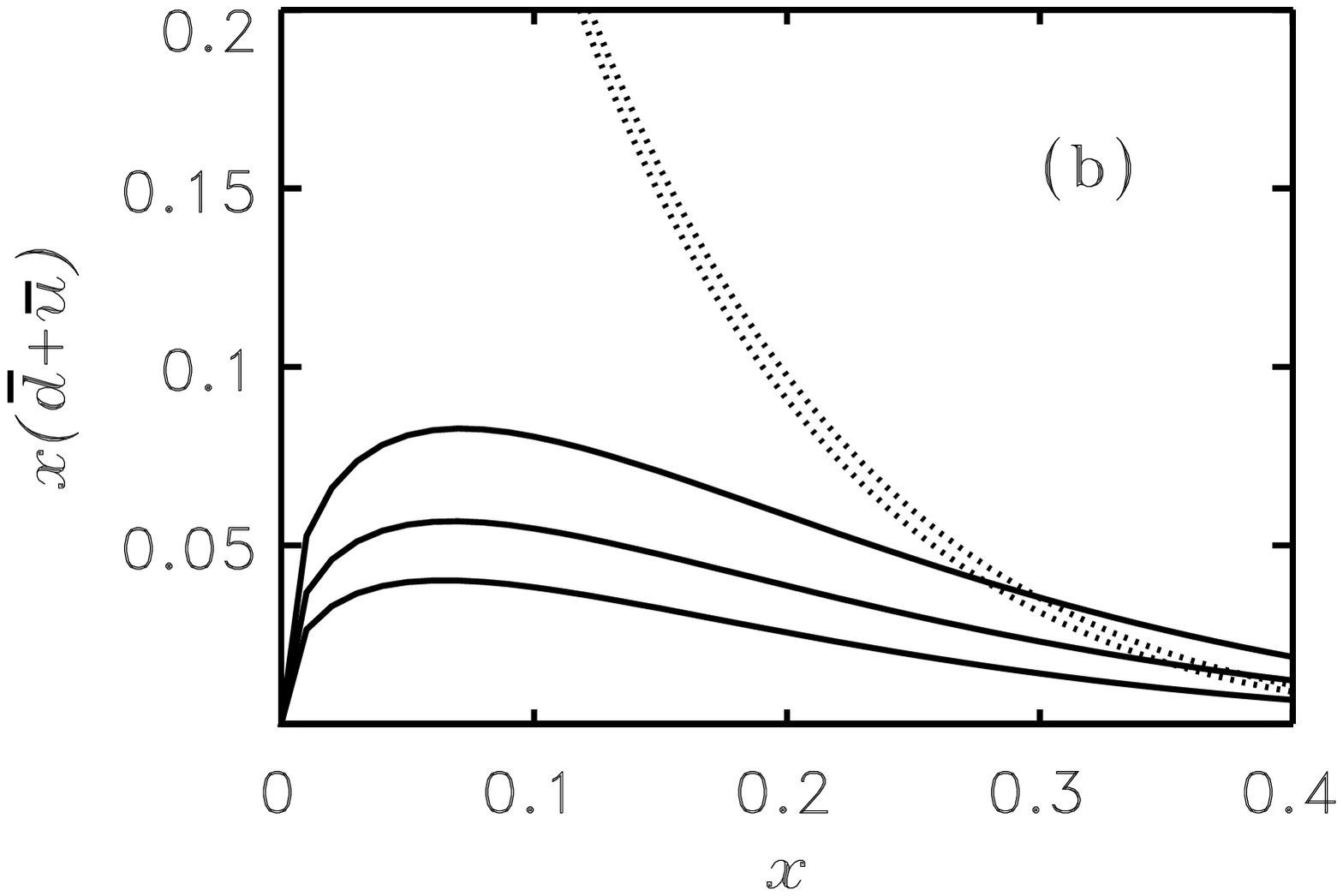,height=1.5in}}
\end{picture}}
\caption{(a) $\pi N$ and $\pi \Delta$ momentum distribution functions,
	with dipole form factor cut-offs $\Lambda_{\pi N}=1$~GeV and
	$\Lambda_{\pi \Delta}=1.5$~GeV.
	(b) Contribution from pion sea to total $\bar u + \bar d$, with
	$\Lambda_{\pi N} = \Lambda_{\pi\Delta} = 1$~GeV
	(largest curve),
	$\Lambda_{\pi N} = 1$~GeV, $\Lambda_{\pi\Delta} = 1.5$~GeV
	(middle), and
	$\Lambda_{\pi N} = \Lambda_{\pi\Delta} = 1.5$~GeV
	(smallest).
	The dotted curves are the CTEQ4	\protect\cite{CTEQ4} and
	MRS98 \protect\cite{MRS98} global parameterisations.}
\end{figure}

The contributions to the antiquark distribution in the proton
can be written as convolutions of the meson distribution functions
and the antiquark distribution in the (on-mass-shell) pion:
\begin{eqnarray}
\label{conv}
\delta \bar q(x)
&=& \sum_{B=N,\Delta} \int_x^1 {dy \over y}
	f_{\pi B}(y)\ \bar q^{\pi}(x/y),
\end{eqnarray}
where the antiquark distribution in the pion is taken from $\pi N$
Drell-Yan experiments.
The contribution to the total $\bar u$ and $\bar d$ distributions
from the non-perturbative cloud is shown in Fig.1(b), compared with
the CTEQ4 \cite{CTEQ4} and MRS98 \cite{MRS98} parameterisations.
While at small $x$ the calculated distributions lie safely below
the parameterisation, at large $x$ the pion cloud already saturates
the total sea with cut-offs
$\Lambda_{\pi N} \sim 1$~GeV, $\Lambda_{\pi\Delta} \sim 1.5$~GeV
--- although one should add a cautionary note that the antiquark
distribution at large $x$ is not determined very precisely.

Because the meson cloud model is at most a model of part of
the non-perturbative sea, it can only be reliably applied to
describing the non-singlet $\bar d - \bar u$ distribution.
To reconstruct the ratio $\bar d/\bar u$ from the calculated
difference we use, following E866, the total $\bar d + \bar u$
from the CTEQ4 parameterization \cite{CTEQ4} as input.
With the above form factor cut-offs, one can get a good fit to the
large-$x$ data \cite{DYN} stemming from the cancellation of some
of the $\bar d$ excess by the $\pi\Delta$ component.
On the other hand, at $x < 0.2$ the asymmetry is now underestimated
somewhat with the hard $\pi\Delta$ component \cite{DYN}.
This suggests that there may be room for other mechanisms which
could account for the missing strength.

Going beyond explanations involving meson clouds, one can also
investigate the possibility that the bare nucleon itself could
be asymmetric with respect to $\bar u$ and $\bar d$.
As suggested long ago by Field and Feynman \cite{FF}, the Pauli  
exclusion principle can contribute to the asymmetry on the basis
of the $u$ and $d$ valence quarks being unequally represented in
the proton, thereby affecting the likelihood with which $q\bar q$ 
pairs can be created in different flavour channels.

In a simple model in which the nucleon is considered to be
composed of 3 quarks in the ground state, insertion of $q\bar q$
pairs split from the incoming virtual photon leads to a ratio of
antiquarks in the proton which must satisfy
$\bar d : \bar u = 5 : 4$.
More quantitative estimates based on the MIT bag model \cite{BAG}
showe that the normalisation, $\Delta^{\rm Pauli}$, of the
$\bar d - \bar u$ difference arising from Pauli blocking could
be as large as 25\%.
Phenomenologically, one can parameterise this contribution as
$(\bar d~-~\bar u)^{\rm Pauli}
= \Delta^{\rm Pauli} (\alpha+1) (1-x)^\alpha$,
where $\alpha$ is some large power.
Because the E866 data implies a softer asymmetry than typical global
fits of total sea quark distributions would give, empirically the
power $\alpha$ should be $> 10$ rather than the 5--7 that has been
common for the total $\bar q$ fits \cite{CTEQ4,MRS98}.

The Pauli effect will produce an excess of $\bar d$ over $\bar u$
over the whole range of $x$, so that it cannot lead to any cancellation
of the large-$x$ asymmetry.
To be consistent with the trend of the large-$x$ data, especially
for the $\bar d/\bar u$ ratio, one needs therefore to keep the
$\pi N N$ contribution softer than the $\pi N \Delta$.
Taking the $\pi N$ and $\pi\Delta$ contributions calculated with   
$\Lambda_{\pi N} = 1$~GeV and $\Lambda_{\pi\Delta} = 1.5$~GeV as
above, we show in Fig.~2 the combined effects of pions and
antisymmetrisation.
For the latter the exponent $\alpha = 14$, and the normalisation is  
$\Delta^{\rm Pauli} \approx 7\%$, which is at the lower end of the   
expected scale but consistent with the bag model calculations \cite{BAG}.
Together with the integrated asymmetry from pions,
$\Delta^{\pi} \sim 0.05$, the combined value  
$\Delta = \Delta^{\pi} + \Delta^{\rm Pauli} \approx 0.12$ is in quite
reasonable agreement with the experimental result, $0.100 \pm 0.018$
from E866 and $0.148 \pm 0.039$ from NMC.

\begin{figure}[ht]
\centering{
\begin{picture}(-30,150)(180,0)
\psfig{figure=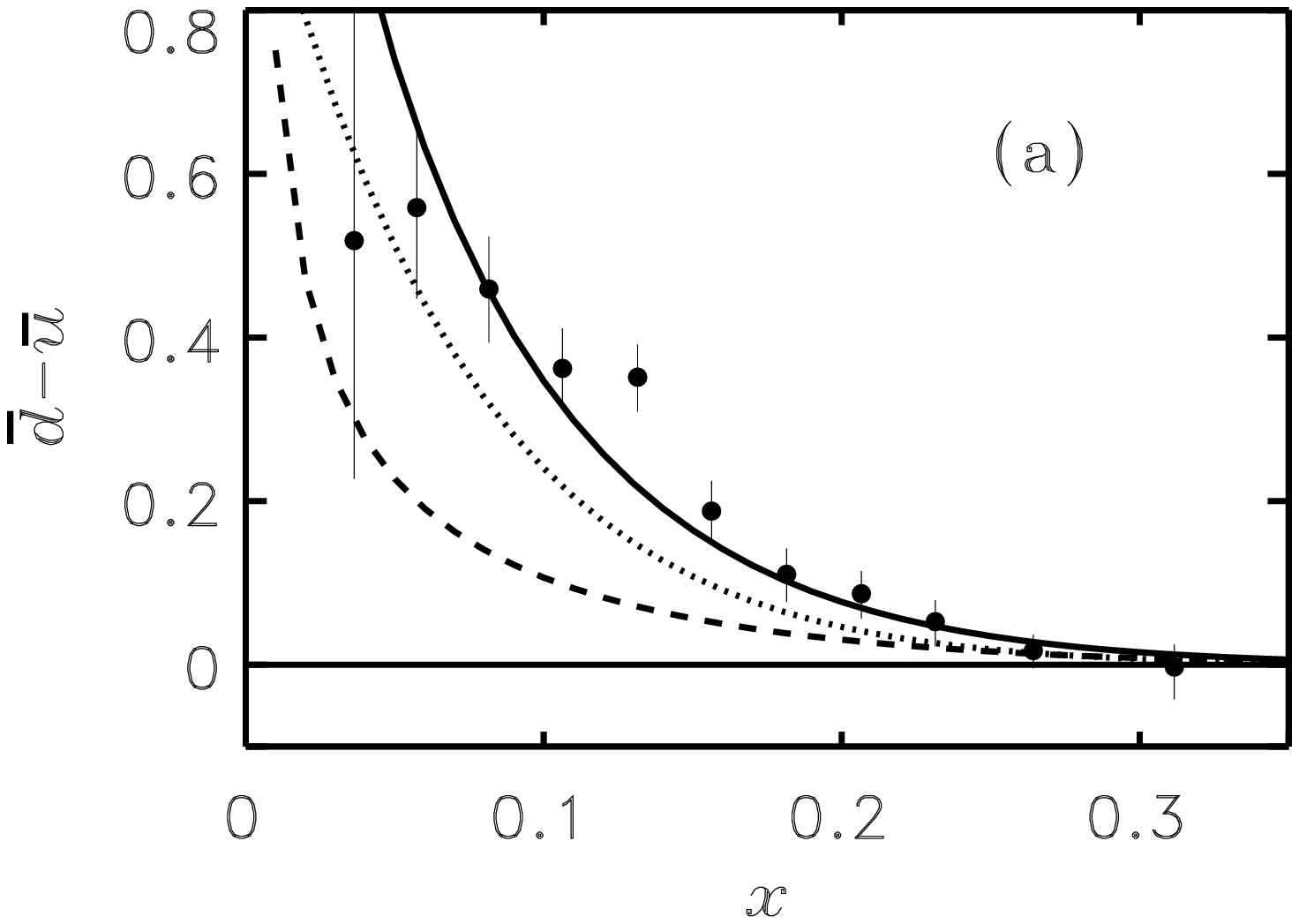,height=1.5in}
\put(0,0){\psfig{figure=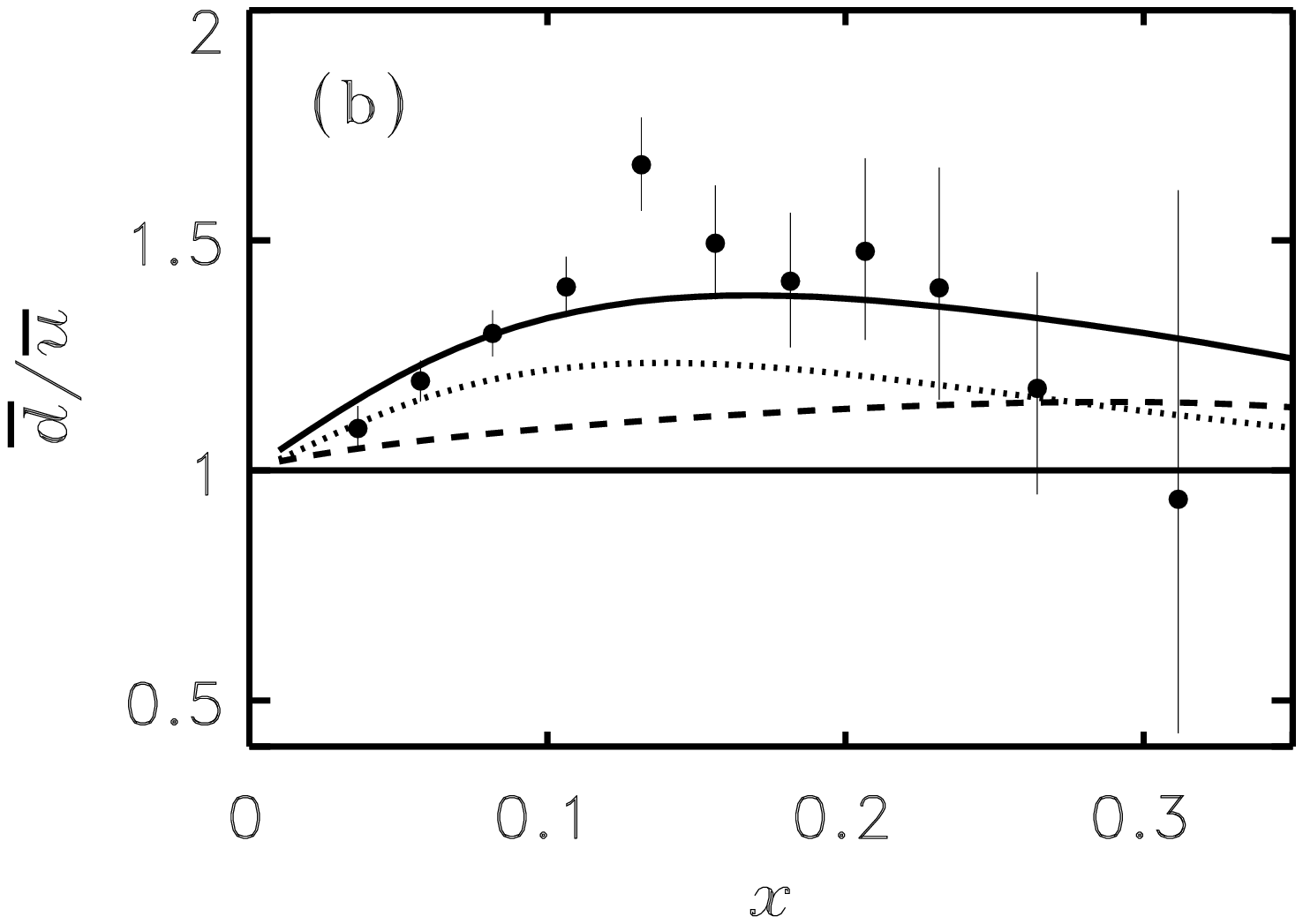,height=1.5in}}
\end{picture}}
\caption{Contributions from pions with $\Lambda_{\pi N} = 1$~GeV
	and $\Lambda_{\pi\Delta} = 1.5$~GeV (dashed) and from
	antisymmetrisation (dotted) to the
	(a) $\bar d - \bar u$ difference and
	(b) $\bar d/\bar u$ ratio,
	and the combined effect (solid).}
\end{figure}

%%%%%%%%%%%%%%%%%%%%%%%%%%%%%%%%%%%%%%%%%%%%%%%%%%%%%%%%%%%%%%%%%%%%%%%%
\section{How Strange is the Nucleon?}

There has been a lot of discussion recently about strange matrix
elements of the nucleon \cite{REP}.
Interest in this subject was largely generated by the deep-inelastic
scattering experiments with polarised targets at CERN and SLAC, which
implied a large polarised strange quark distribution in the proton.
At about the same time a measurement of the elastic neutrino--proton
scattering cross section at lower values of $Q^2$ had also suggested a
non-zero value for the strange axial vector form factor of the proton.

In response to these observations, other processes were sought
in which traces of strangeness in the nucleon could be detected,
such as parity-violating electron scattering.
Some of these have since been performed (at MIT-Bates \cite{SAMPLE}),
while others (at Jefferson Lab) will soon provide valuable data on
the strangeness radius and magnetic moment of the nucleon.

In the DIS regime, the difference between strange and antistrange
quark distributions in the nucleon has recently come to prominence
again with the availability of new $\nu$ and $\bar \nu$ DIS data
from the CCFR collaboration \cite{CCFR}, which were analysed for
a possible non-zero $s$--$\overline s$ difference.
Perturbative QCD alone would be expected to produce identical $s$
and $\overline s$ distributions, while any asymmetry would have to
be non-perturbative in origin.
Such an asymmetry arises naturally in a meson cloud picture of the
nucleon, where the strangeness of the nucleon is assumed to be
carried by the kaon--hyperon components of the physical nucleon.
The $s$ and $\overline s$ quarks therefore have quite different
origins in this model.
Because the $s$ quark originates in the $\Lambda$, its distribution
is like that of the $u$ quark in the proton, roughly $\sim (1-x)^3$
at large $x$.
The $\overline s$ in the kaon, on the other hand, is much harder,
$\sim (1-x)$.
Since the $K\Lambda$ distribution function $f_{K\Lambda}(y)$ is
fairly symmetric around $y=1/2$, upon convoluting the momentum
distribution with the $s^{\Lambda}$ and $\overline s^K$ distributions,
this asymmetry is largely preserved, leading to harder $\bar s$
distributions at large $x$ in the proton.

The asymmetry in the kaon cloud model turns out to be very small,
and broadly consistent with the CCFR experiment \cite{CCFR} within
the given errors for not too large values of $\Lambda_{K\Lambda}$,
Fig.~3(a).
To obtain the difference $s$--$\overline s$ we have used the
absolute values of $s + \overline s$ from the parameterisations
of Refs.\cite{CTEQ4,MRS98}.
We should point out, however, that there exists some controversy
regarding the overall normalisation of the deep-inelastic neutrino
data from which the strange quark distribution was extracted,
resulting from an apparent inconsistency between the neutrino data
and data on inclusive charm production \cite{GRV,BLT}.
In addition, the CCFR data were collected with Fe nuclei targets,
so that one needs to consider possible nuclear EMC corrections in
the data analysis \cite{BLT} before making any definitive conclusions
about $s$ and $\overline s$.

\begin{figure}[ht]
\centering{
\begin{picture}(-30,150)(180,0)
\psfig{figure=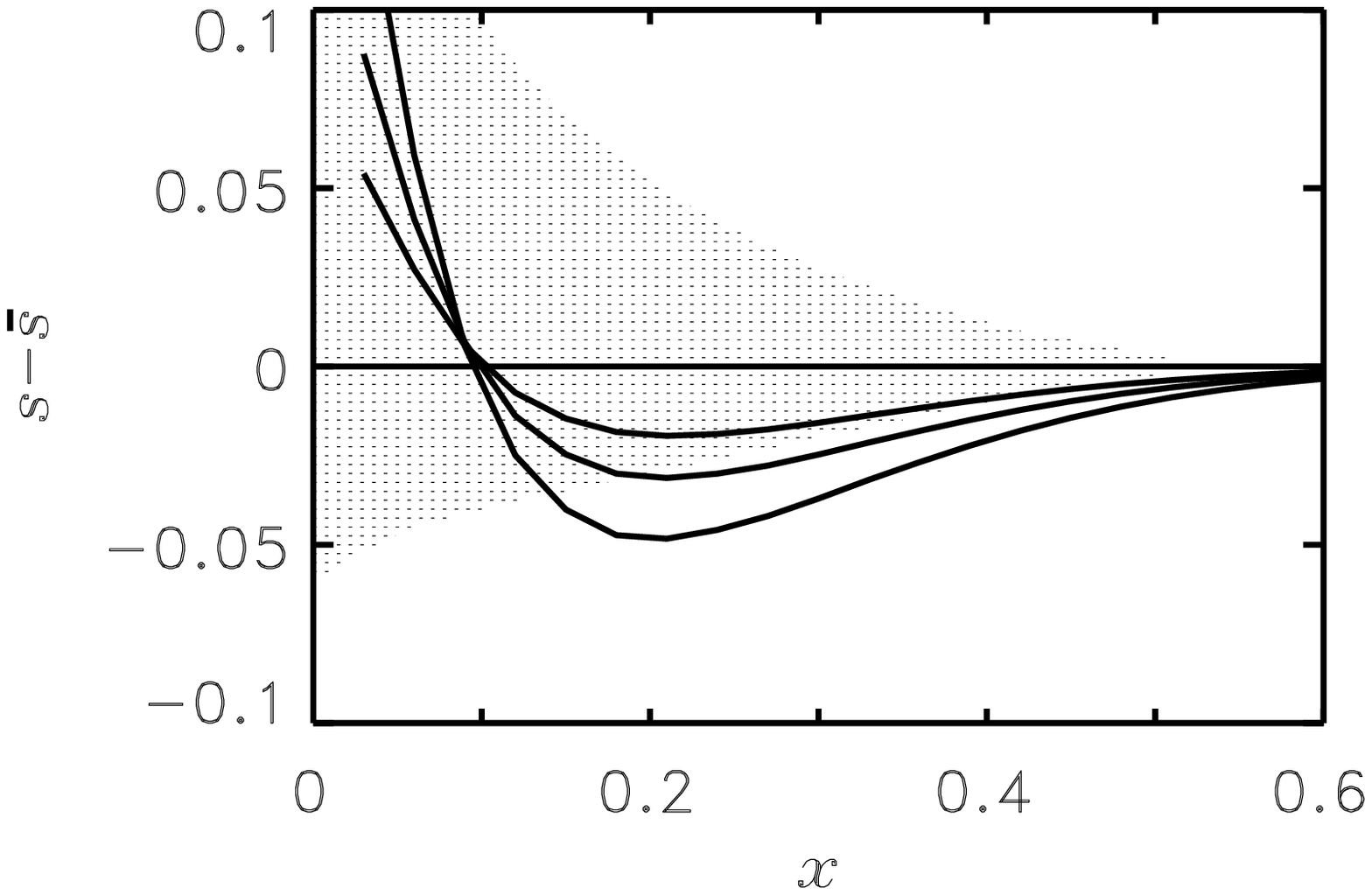,height=1.5in}
\put(0,0){\psfig{figure=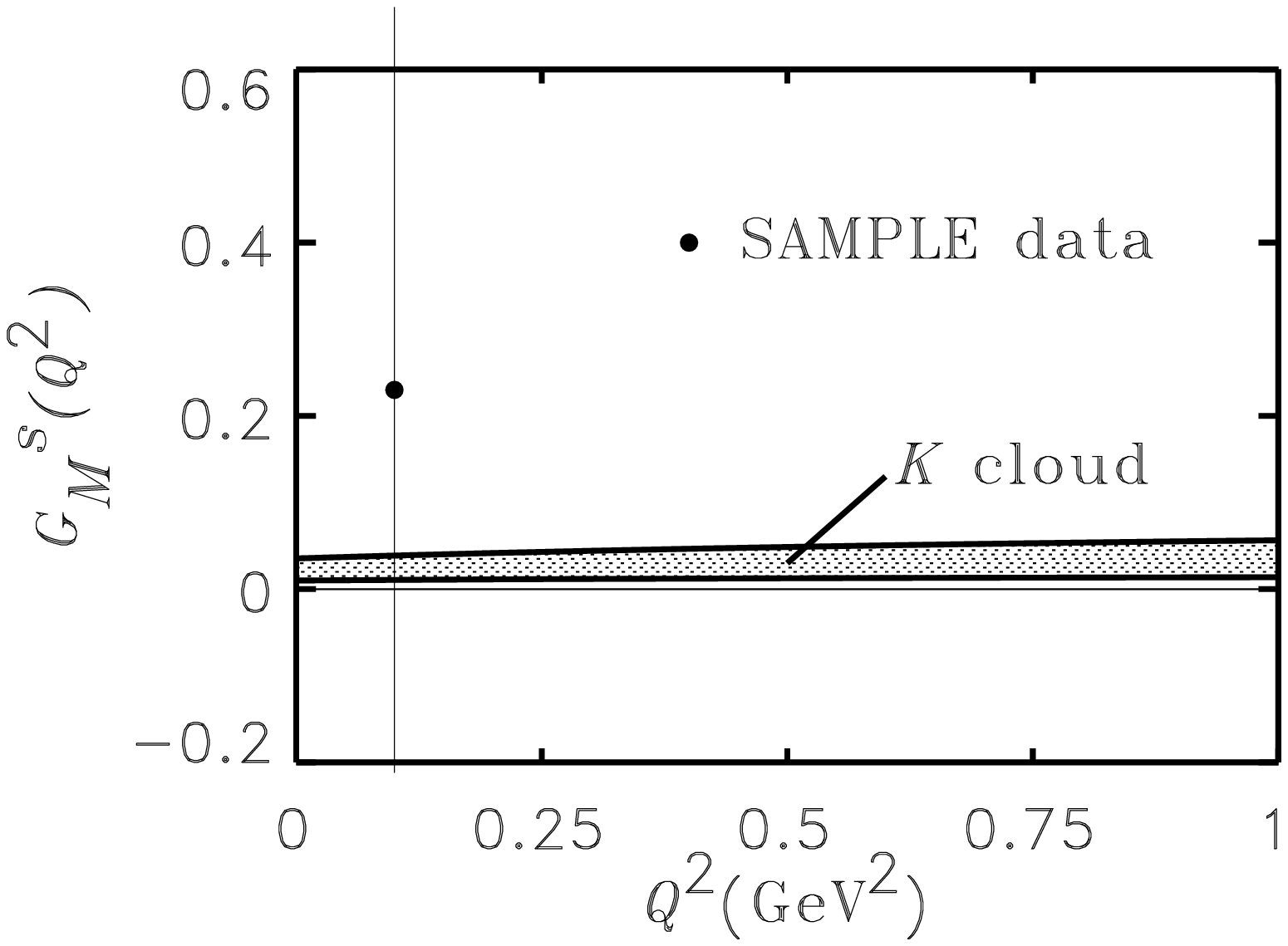,height=1.5in}}
\end{picture}}
\caption{(a) Strange--antistrange quark distribution asymmetry in the
	nucleon.
	The solid lines correspond to the asymmetry calculated for
	$\Lambda_{K\Lambda} = 0.7$ GeV (smallest asymmetry), 1 GeV and
	1.3 GeV (largest asymmetry), while the shaded region represents
	the uncertainty range of the data \protect\cite{CCFR}.
	(b) Strange magnetic form factor of the proton $G_M^S(Q^2)$
	as a function of $Q^2$.
	The shaded region is the kaon cloud prediction, for
	$\Lambda_{K\Lambda}$ = 1 (lower curve) and 3 (upper curve) GeV.
	The data point is from the SAMPLE experiment \protect\cite{SAMPLE}.}
\end{figure}

Within the same formalism one can also calculate the strangeness form
factors of the nucleon at low $Q^2$.
The strangeness (Sachs) radius (defined in terms of the strange electric
form factor) is found to be very small and negative, in the vicinity
$r_s^2 \approx -0.004 \rightarrow -0.008$ fm$^2$ for $KN\Lambda$
vertex function cut-offs of $\Lambda_{K\Lambda}=0.7$--$1.3$ GeV.
The strangeness magnetic form factor, on the other hand, suffers
from spurious contributions arising from the breaking of Lorentz
covariance due to the use of the impulse approximation on the
light-cone (where the calculation is performed \cite{MM}).
Once these are removed, according to the prescription outlined in
Ref.\cite{KF}, the strange magnetic form factor turns out to be
small and positive, as in Fig.~3(b), consistent with the trend
of the SAMPLE data \cite{SAMPLE}, and largely independent of the
details of the $KN\Lambda$ vertex function.

%%%%%%%%%%%%%%%%%%%%%%%%%%%%%%%%%%%%%%%%%%%%%%%%%%%%%%%%%%%%%%%%%%%%%%%%
\section{Intrinsic Charm}

In 1997 the H1 \cite{H1} and ZEUS \cite{ZEUS} Collaborations at HERA
announced an excess of events at large $x$ and $Q^2$ in $e^+ p$ neutral
current (NC) and charged current (CC) deep-inelastic scattering,
which prompted numerous speculations about whether one has seen
evidence for physics beyond the standard model, such as leptoquarks
or contact interactions.
Since then the apparent signal has decreased, although an excess still
persists.
More interesting from the point of view of the physics of
non-perturbative QCD discussed here, the HERA data opened
up an avenue through which to investigate the structure of
the heavy quark sea.

It was suggested \cite{KLT} that a large enhancement of the cross
sections could be achieved by slightly increasing some of the quark
distributions, such as the $u$ quark, at large $x$.
Unfortunately, as pointed out in Ref.\cite{RB}, the size of the
additional $u$ quark contribution necessary to achieve sufficient
enhancement would significantly overestimate the SLAC large $x$ data.
On the other hand, the possibility that the charm quark distribution
might be enhanced at large $x$, thereby producing a similar effect,
has not been ruled out, and indeed, one could more easily imagine
that the charm quark distribution could be rather hard, owing to
the large $c$ quark mass.

The effect of a non-perturbative, or intrinsic, charm component on
the DIS cross sections at HERA kinematics was recently investigated
by Gunion and Vogt \cite{GV} within a model of the 5-quark component
of the nucleon wave function on the light-cone \cite{BROD}, with
normalisation fixed to 1\% \cite{GV}.
After evolving to the HERA kinematics, this intrinsic charm distribution,
while considerably harder than that generated through pQCD, was still too
soft to account for the excess HERA events \cite{GV}.
As an alternative to this intrinsic charm model, the charmed sea
was taken in Refs.\cite{PNNDB} to arise from the quantum fluctuation
of the nucleon to a virtual $D^- + \Lambda_c$ configuration, along
the lines of the $\pi$ and $K$ cloud models discussed above.
A natural prediction of this model is non-symmetric $c$
and $\overline c$ distributions.

Because of the large mass of the $c$ quark, one can approximate the
$\overline c$ distribution in the $D^-$ meson \cite{PNNDB} and the
$c$ distribution in the $\Lambda_c^+$ by:\ 
$\overline c^{D^-}(x) \approx \delta(x-1)$,\
$c^{\Lambda_c^+}(x)\ \approx\ \delta(x-2/3)$.
Quite interestingly, the shape of the resulting $\delta c$ quark
distributions is quite similar to that in the intrinsic charm model
of Refs.\cite{GV,BROD}.
However, although the model of \cite{GV,BROD} assumes identical
shapes for the non-perturbative $c$ and $\overline c$ distributions,
the meson cloud gives a significantly harder $\overline c$ distribution.

\begin{figure}[ht]
\centering{
\begin{picture}(-30,150)(180,0)
\psfig{figure=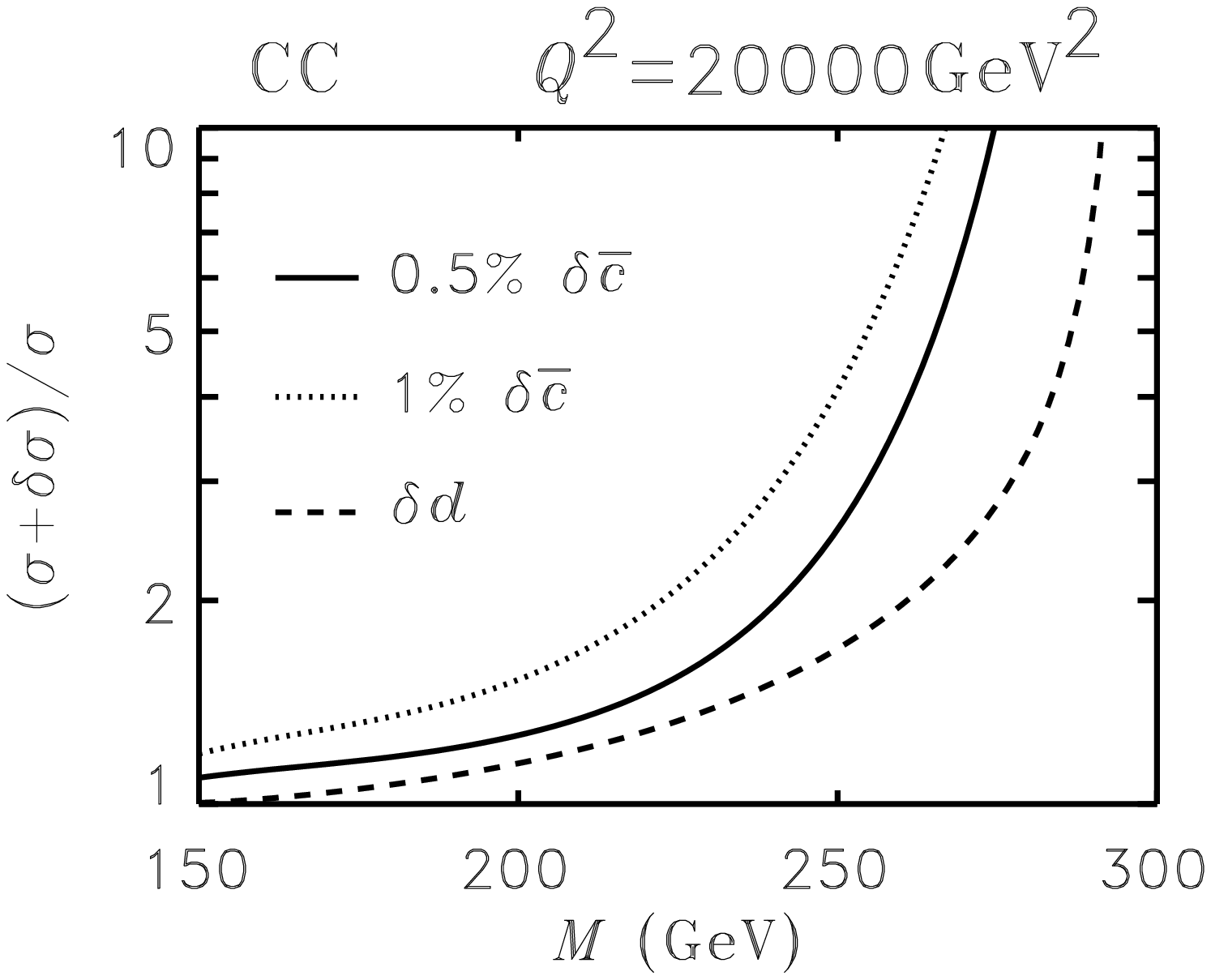,height=1.5in}
\put(0,0){\psfig{figure=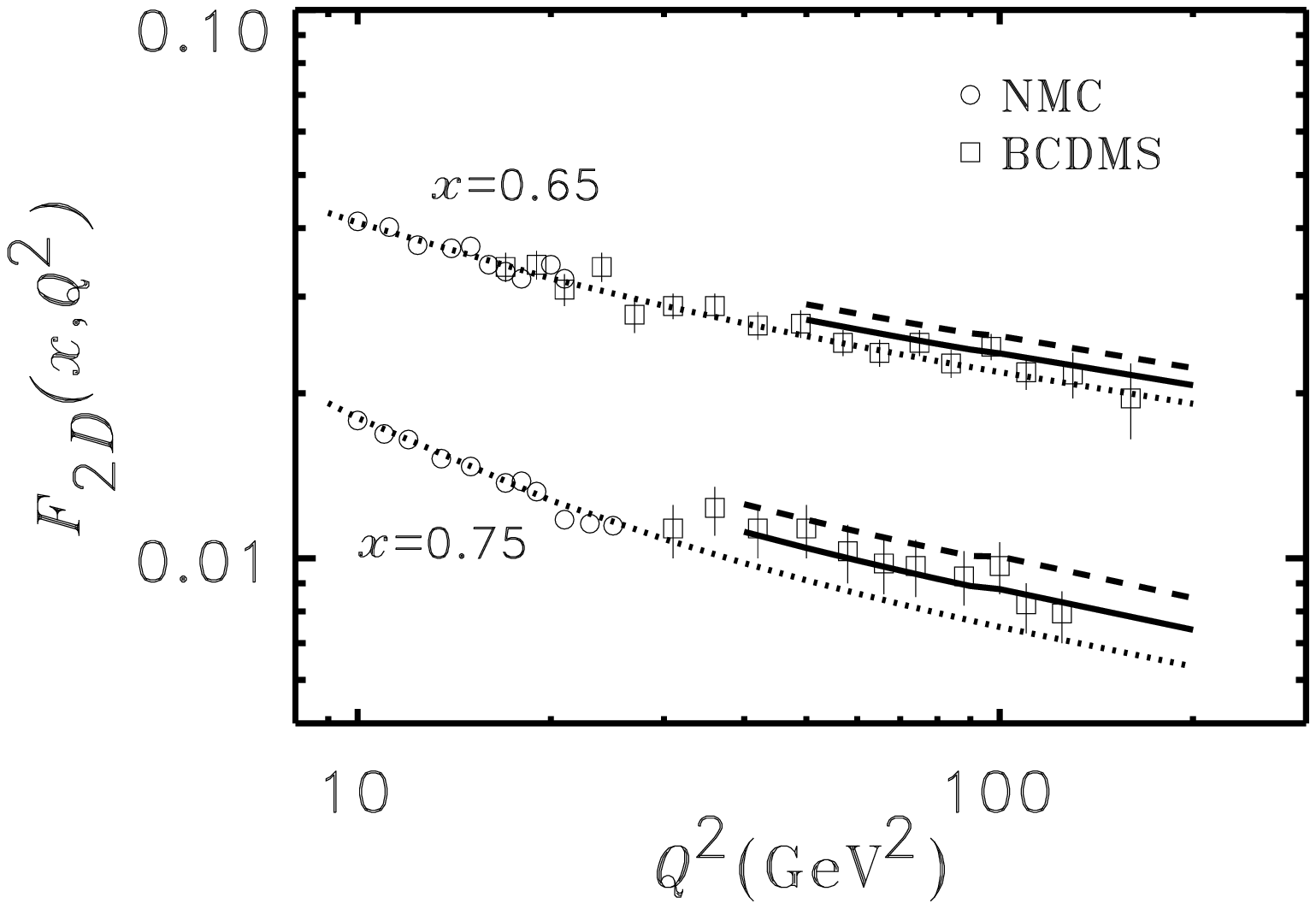,height=1.5in}}
\end{picture}}
\caption{(a) Ratio of modified to standard DIS model CC cross sections
	at $Q^2=20000$~GeV$^2$, with the modifications arising from
	0.5\% and 1\% additional $\delta \overline c$ distributions, 
	as well as a modified $d$ quark distribution at large $x$.
	(b) Structure function of the deuteron, $F_{2D}$, with 0.5\%
	(solid) and 1\% (dashed) additions of non-perturbative charm
	to the global fit (dotted) from NMC.}
\end{figure}

Evolving the charm and anticharm distributions to an average value
of $Q^2 = 20000$~GeV$^2$, the effects on the CC cross section are
seen in Fig.4(a).
With a 0.5\% (1\%) intrinsic charm component the CC cross section
increases by a factor $\sim$ 2 (3) for $200 < M < 250$~GeV, which
is similar to the excess observed by H1 \cite{H1} in this region.

While the values of $W^2$ corresponding to the $x$ and $Q^2$ values
are too low for the SLAC data \cite{RB} to be sensitive to the charm
component of the nucleon wave function, data are, however, available
from the BCDMS Collaboration in the region $x > 0.6$ which are above
charm threshold \cite{BCDMS}, and can be used to provide limits on
the size of the intrinsic charm.
The effect of adding the intrinsic $\overline c$ distribution to the
data on the deuteron structure function is illustrated in Fig.4(b).
With the addition of the 0.5\% contribution there does not seem to be
any inconsistency with the data \cite{BCDMS} and, indeed, the agreement
is slightly improved at $x=0.75$.
On the other hand, the 1\% case may be a little too high for comfort.

Another possibility which could lead to additional enhancement of the
cross sections at large $x$ would be a larger $d$ quark distribution.
The recent reanalysis, for example, of the SLAC deuteron data in
Ref.\cite{NP} suggested that the valence $d/u$ ratio does not tend
to zero as $x \rightarrow 1$, as assumed in global fits to data
\cite{CTEQ4}, but rather is consistent with the expectation of
perturbative QCD, namely $d/u \rightarrow 1/5$ \cite{FJ}.
The effect of the modified $d$ quark distribution is comparable to
that of the 0.5\% anticharm scenario, as seen in Fig.4(a).
The fact that all standard sets of parton distributions assume that
$d/u \rightarrow 0$ as $x \rightarrow 1$ means that there is a possible
source of systematic error in the modeling of ``background'' events  
which should be accounted for.

%%%%%%%%%%%%%%%%%%%%%%%%%%%%%%%%%%%%%%%%%%%%%%%%%%%%%%%%%%%%%%%%%%%%%%%%
\section{Conclusion}

The high precision deep-inelastic, Drell-Yan and other experiments
in the last few years have only begun to unravel the rich substructure
of the proton sea.

We have, for the first time, at our disposal important new data from
the E866 Collaboration which map out the $x$-dependence of the
$\bar d / \bar u$ asymmetry in the proton.
Most importantly, the E866 results confirm the recent observations
that the $\bar d$ and $\bar u$ content of the proton is not symmetric,
while an interesting new feature of the data is the relatively fast
downturn in the $\bar d/\bar u$ ratio beyond $x \sim 0.15$,
which drops rapidly back to unity by $x \sim 0.3$.
The current evidence from large $x$ clearly indicates the necessity
of a $\pi\Delta$ component in the nucleon wave function, one which
is harder in momentum space than the $\pi N$ component.
Consistency with data for the sum of $\bar d$ and $\bar u$ at $x > 0.2$
requires that both the $\pi N N$ and $\pi N \Delta$ form factors be
relatively soft, making it difficult to avoid underestimating the E866
asymmetry at intermediate $x$, and leaving room for other effects,
such as the Pauli exclusion principle, to make up the difference.
Along the lines of previous estimates of the Pauli effect,
we find the contribution to the $\bar d-\bar u$ difference from
antisymmetrisation to be significant in magnitude, and particularly  
important small at $x$.
Our results suggest that the best description of the E866 data is
that in which chiral symmetry and antisymmetrisation play roughly
equal roles --- consistent with the findings of the earlier analysis
\cite{MTS} of the NMC data for $F_2^p - F_2^n$.
We note, however, that it would be helpful to have more data at
large $x$, where the error bars are largest, to verify the downward
trend of $\bar d - \bar u$, and to further explore the possible
discrepancy between the Drell-Yan and NMC data.

Using the same meson cloud framework, one can estimate the asymmetry
between the $s$ and $\overline s$ quark distributions in the nucleon,
which has been studied by the CCFR Collaboration in $\nu$ charm
production \cite{CCFR}.
The magnitude of the $s$--$\overline s$ difference turns out to be
very small, consistent with current experimental errors, with the
$\overline s$ distribution slightly harder than the $s$.
More definitive conclusions will be reached with more statistics
on the charm production data, and the apparent discrepancy between
the inclusive deep-inelastic muon and neutrino data and those on
$c\bar c$ production resolved \cite{CCFR,GRV,BLT}.

Finally, the task of identifying a possible intrinsic charm
component of the nucleon can be advanced by tagging charm final
states in $J/\psi$ production at HERA to measure the charm
structure function, or through a thorough comparison of NC and CC
cross sections, for both $e^+ p$ and $e^- p$ collisions, to enable
one to determine whether the large-$x$ enhancement of the cross
sections is due to charm or other flavours.
Needless to say, we eagerly await further results from the H1 and ZEUS
Collaborations.
Any experimental evidence supporting the suggestion that the intrinsic
charm could have a strong asymmetry would mean a revision of current
wisdom, and would undoubtedly lead us to a deeper understanding of the
structure of hadronic systems.

%%%%%%%%%%%%%%%%%%%%%%%%%%%%%%%%%%%%%%%%%%%%%%%%%%%%%%%%%%%%%%%%%%%%%%%%
\section*{Acknowledgments}

I would like to thank M. Malheiro, J. Speth, F.M. Steffens and
A.W. Thomas for their contributions to my understanding of the
material presented here, and the Special Research Centre for the
Subatomic Structure of Matter for hospitality and support during
this workshop.

%%%%%%%%%%%%%%%%%%%%%%%%%%%%%%%%%%%%%%%%%%%%%%%%%%%%%%%%%%%%%%%%%%%%%%%%

\end{document}